\documentclass{article}
\usepackage{ijcai20}

\usepackage{times}

\usepackage{soul}
\usepackage{url}
\usepackage[hidelinks]{hyperref}
\usepackage[utf8]{inputenc}
\usepackage[small]{caption}
\usepackage{graphicx}
\usepackage{amsmath}
\usepackage{booktabs}
\usepackage[ruled,linesnumbered]{algorithm2e}
\usepackage{algorithmic}
\newtheorem{observation}{Observation}
\usepackage{xcolor}

\title{FedCoin: A Peer-to-Peer Payment System for Federated Learning}
\author{
	Yuan Liu$^1$\footnote{Contact Author}\and
	Shuai Sun$^1$\and
	Zhengpeng Ai$^{1}$\And
	Shuangfeng Zhang$^1$\and
	Zelei Liu$^{2}$\and
	Han Yu$^{2}$ 
	\\
	\affiliations
	$^1$Northeastern University, Shenyang, China\\
	$^2$Nanyang Technological University, Singapore\\
	\emails
	liuyuan@swc.neu.edu.cn, sunshuai.edu@gmail, im@aizhengpeng.cn, 1871161@stu.neu.edu.cn\\zelei.liu@ntu.edu.sg, han.yu@ntu.edu.sg
}

\begin{document}

\maketitle

\begin{abstract}
Federated learning (FL) is an emerging collaborative machine learning method to train models on distributed datasets with privacy concerns. To properly incentivize data owners to contribute their efforts, Shapley Value (SV) is often adopted to fairly assess their contribution. However, the calculation of SV is time-consuming and computationally costly. In this paper, we propose FedCoin, a blockchain-based peer-to-peer payment system for FL to enable a feasible SV based profit distribution. In FedCoin, blockchain consensus entities calculate SVs and a new block is created based on the proof of Shapley (PoSap) protocol. It is in contrast to the popular BitCoin network where consensus entities ``mine'' new blocks by solving meaningless puzzles. Based on the computed SVs, a scheme for dividing the incentive payoffs among FL clients with nonrepudiation and tamper-resistance properties is proposed. Experimental results based on real-world data show that FedCoin can promote high-quality data from FL clients through accurately computing SVs with an upper bound on the computational resources required for reaching consensus. It opens opportunities for non-data owners to play a role in FL.
\end{abstract}

\section{Introduction}
Nowadays, many businesses generate large amount of data through usage, and rely on making informed decisions using machine learning (ML) based on such data in order to thrive. With changing regulatory scene, ML is facing increasingly difficult challenges with respect to the usage of such data. Data are often collected and owned by multiple distributed entities. They often contain sensitive or private information and is cannot be stored in a centralized server without violating privacy protection laws. In recent years, federated learning (FL) has emerged as a promising solution to these challenges \cite{yang2019federated}.  

In FL, each entity trains its local model and contributes the local model parameter updates to a center server to help build a more powerful global FL model. Compared with centralized ML methods, FL not only reduces communication costs by transmitting model updates instead of raw data, but also reduce the computational costs of the server by leveraging computing power from the clients. Moreover, since local data never leave the data owners, FL improves user privacy \cite{DeepFLModel2017,FLScalSystemDesign2019}.

From the above description, it is clear that FL clients are making significant contributions towards the FL model. Thus, in order to sustain an FL community, it is important for FL clients to be properly incentivized. For this to happen, FL clients must be treated fairly \cite{yu2019}. Existing FL incentive schemes generally agree that fair treatment of FL clients shall be based on a fair assessment of their contributions to the FL model \cite{FLSurvey2019}. Currently, the most widely adopted method for fair assessment of FL client contribution is that of Shapley Values (SVs) \cite{EfficientDataSV2019,Song-et-al:2019}.


SV is a popular notion in fairly distributing profits earned by a coalition among its contributors. It has been applied in various fields, ranging from economics, information theory, and ML. The reason for its broad application is that the SV divides the payoff with attractive properties such as fairness, individual rationality, and additivity. However, SV based distribution solution often takes exponential time to compute with a complexity of $\mathcal{O}(n!)$ where $n$ is the number of data items. Even though the computational complexity can be reduced through approximating SV with marginal error guarantees \cite{EfficientDataSV2019}, it is still computationally costly. 


In order to help FL systems compute SVs to support sustainable incentive schemes, we propose a blockchain-based peer-to-peer payment system \textit{FedCoin}. The Shapley value of each FL client, reflecting its contribution to the global FL model in a fair way, is calculated by the Proof of Shapley (PoSap) consensus protocol which replaces the traditional hash-based protocol in existing proof of work(PoW) based blockchain systems. All the payments are recorded in the block in an immutable manner. Under FedCoin, paying out incentives to FL clients does not need to rely on a central FL server. Based on this, FedCoin provides a decentralized payment scheme for FL so that incentives for all participants can be delivered in third-party-free manner with nonrepudiation and tamper-resistance properties. 

Extensive experiments based on real-world data show that FedCoin is able to properly determine FL clients' Shapley Value-based contribution to the global FL model with an upper bound on the computational resources required for reaching consensus. To the best of our knowledge, FedCoin is the first attempt to leverage blockchain technology in federated learning incentive scheme research. It opens up new opportunities for entities which has computational resources but without local data to contribute to federated learning.

\section{Related Work}  
The incentive mechanism design is an important research direction in the field of federated learning \cite{FLSurvey2019,yang2019federated}. In \cite{KangXNXZ19}, the contract theory is employed to improve the accuracy of model training considering the unreliable data contributors. A consortium blockchain architecture is applied to build a decentralized reputation system. In \cite{FLResourceOandIM2019}, a Stackelberg-game based incentive mechanism is designed to optimize the utilities of both FL clients and the FL server. These works focus on optimizing the rewards for self-interested FL clients and FL customers who pay to use the FL model. Our study is compatible with these works in terms of determining the payment budget for the FL customers.

In field of ML, SV has also be studied widely for various purpose. SV can be applied in feature selection, ranking the importance of training data, which is further applied in explaining the behavior of ML models \cite{DBLP:journals/corr/abs-1909-06143}. Since the computation complexity is $\mathcal{O}(n!)$, approximations of SV also attract many attentions. In \cite{DBLP:conf/icml/AnconaOG19}, a polynomial-time approximation of SV is proposed for deep neural network. Group sampling based approximation is studied in \cite{EfficientDataSV2019}. In this work, our objective is not to decrease the computational complexity, but to establish a scheme so that distributed computational resources, which are otherwise wasted, can be leveraged to help FL systems calculate SVs.  

Blockchain has been widely applied in addressing the security problems in FL applications \cite{dillenberger2019blockchain,FLSurvey2019,yang2019federated}. FLChain \cite{DBLP:conf/bigcom/BaoSXHH19} and BlockFL \cite{kim2019blockchained} have been proposed to record the local model parameter updates in the temper-resistant manner. A blockchain-based FL was proposed in
\cite{ramanan2019baffle} so as to remove the need for an FL server. A blockchain-based trust management system was proposed in \cite{KangXNXZ19} to assist FL server to select reliable and high quality data owners as FL clients. These blockchain systems are used as immutable ledgers to record local gradients and aggregate them in a trusted manner. Our work will adopt the blockchain network as a computational engine and payment distribution ledger, which is the first of its kind in the current literature.

\section{Preliminaries}
For a typical FL scenario, we take $\mathcal{F}_i(w)=\ell(x_i,y_i;w_t)$ as the loss of prediction on a sample ($x_i$,$y_i$) with model parameters $w$ at the $t$-th round. The parameters $w^t$ is a $d$-dimensional vector. We assume that there are $K$ clients, and each client has a local data set $\mathcal{D}_k$ with $n_k=|\mathcal{D}_k|$. The overall dataset is $\mathcal{D}=\{\mathcal{D}_1,\ldots,\mathcal{D}_K\}$ with $n=|\mathcal{D}|=\sum_{k=1}^Kn_k$.  The objective function to be optimized is:
\begin{equation}
\min_{w\in\mathcal{R}^d}\mathcal{F}(w) \text{\hspace{5mm}where \hspace{5mm}} \mathcal{F}(w)=\frac{1}{n}\sum_{k=1}^K\sum_{i\in\mathcal{D}_k}\mathcal{F}_i(w)
\end{equation}

This optimization problem is generally solved by stochastic gradient descent (SGD)~\cite{goodfellow2016deep} based methods. For example, based on the current model $w_t$, the federated averaging algorithm~\cite{DeepFLModel2017} computes the average gradient $g_k^t=\frac{1}{n_k}\sum_{i\in\mathcal{D}_k}\nabla \mathcal{F}_i(w^t)$ on the local data of client $k$. Each client updates its local model $w_k^{t+1}\leftarrow w_t-\eta g_k^t$, and the FL server aggregates the local models as the global FL model.
\begin{equation}
w_{t+1}\leftarrow \mathcal{A}(\{w_{t+1}^k|k=1,\ldots,K\})
\end{equation}where $\mathcal{A}$ is an aggregation function.
\section{FedCoin}
\begin{figure}[!ht]
	\centering
	\includegraphics[trim = 68mm 22mm 20mm 45mm,  clip, width=1.0\linewidth]{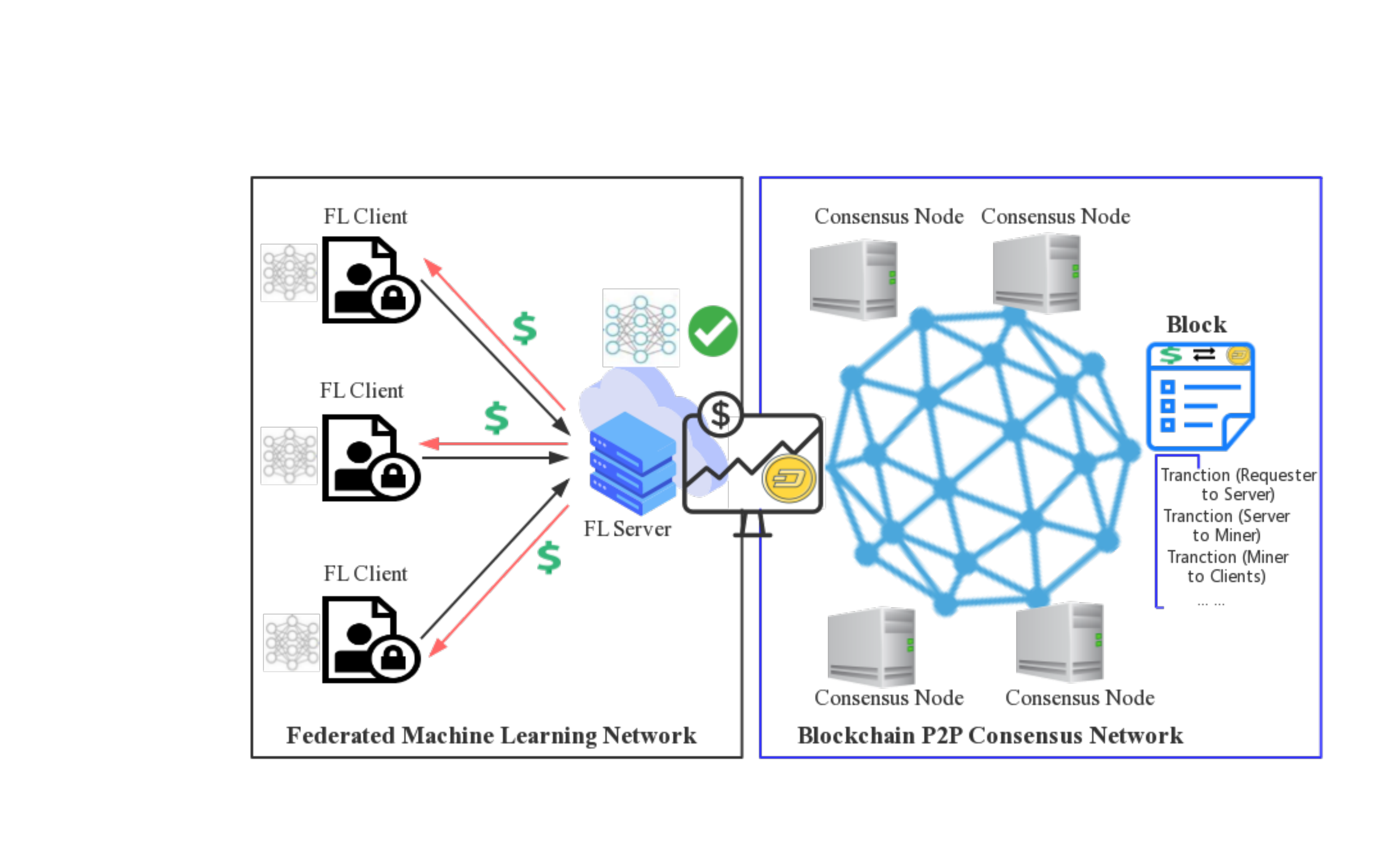}
	\caption{Overview of the Proposed Model}
	\label{FigSystemOverview}
\end{figure}

There are two networks of participants in our system: 1) the FL network, and 2) the peer-to-peer blockchain network (Figure \ref{FigSystemOverview}). 
A \textit{FL model requester} or FL training task requester refers to the entities who need to train an FL network and with a budget of $V$. In the FL network, there is a centralized server, referred as \textit{FL server}, in coordinating the executing of model training and receiving payment $V$ from FL model requester. 

The distributed data owners, called as \textit{FL clients}, participate in a collaborative training task and receive a payment $V$. Each FL client trains its local model and submits the parameter updates to the FL server. The FL server plays three roles. Firstly, it publishes a training task to FL clients with price TrainPrice. Secondly, it aggregates local updates through a secure aggregation protocol \cite{SecureAggragatorCCS2017} and earns a computation payment (ComPrice). Thirdly, it transfer a processing fee SapPrice to the blockchain network to enlist its members' help in calculating the FL model. The total payment of the task (TrainPrice+ComPrice+SapPrice) should be not greater than $V$ in order to sustain payment balance without relying on external transfer of values into this system.

After each global model update, the FL server publishes a task to calculate the contribution by each FL client. The consensus nodes in blockchain network then collaboratively calculate SVs, and the block winner receives a payment of TrainPrice+SapPrice. The winner then divides ComPrice to FL clients according to their respective SVs by creating transactions in the blockchain. In our current design, we only reward clients with postive contributions, but refrain from penalizing clients with negative contributions. All the transactions are recorded in the new block and further updated to the chain.



Therefore, the connection between FL network and blockchain network is a special type of task. A task includes the received local update set $W=\{w_k|k=1,\ldots,K\}$, the aggregation function $\mathcal{A}$, the loss function $\mathcal{F}(w)$, and values SapPrice and TrainPrice for each update round. Note that SapPrice and TrainPrice decreases as the number of training rounds increases, and the total payment for training can be divided among the rounds equally or not. Without loss of generality, the following description focuses on a single training round.


\subsection{Shapley Value Based Blockchain Consensus} 
Upon receiving a Shapley value calculation task from the FL network, the miners in the blockchain network are to calculate the SV vector $S=[s_k]_{k\in[1,K]}$ where $s_k$ is the SV of the client in providing $w_k\in W$. Each miner independently calculates the SV vector following Algorithm \ref{AlgShapley}. Since the objective of the mining process is to competitively calculate SV vectors so as to prove the miner's computation power, we name the algorithm as ``Proof of Shapley (PoSap)''. The input of Algorithm \ref{AlgShapley} comes from the task specifications from the FL network. The output is a new generated block.

\begin{algorithm}[!htb]
\caption{Proof of Shapley (PoSap)}
\label{AlgShapley}
\KwIn{$\mathcal{F}$: Loss function;\\
	\hspace{1.1cm}$\mathcal{A}$: Aggregation function;\\
	\hspace{1.1cm}$W$: Contribution of FL clients in size $K$;\\
	\hspace{1.1cm}$D$: Difficulty in Mining;\\
}
\KwOut{$Blk$: a new block}
Initialize $S=[s_k=0|k=1,\ldots,K]$;\\
time=0;\\
\While{No received $Blk$ OR !VerifyBlock($Blk$)}
{
$S_{t}=[s_k=0|k=1,\ldots,K]$\% temporary store $S$\\
Random generate a rank $R=[r_k|k=1,\ldots,K]$;\\
$S_{t}(R(1))=\mathcal{F}(\mathcal{A}(W(R(1))))$;\\
\For{$i$ from $2$ to $K$}
{
	$S_{t}(R(i))=\mathcal{F}(\mathcal{A}(W(R(1:i))))$;\\
	$S_{t}(R(i))=S_{t}(R(i))-\sum_{j=1}^{i-1}S_{t}(R(j))$;\\
}
$S=\frac{S\times time+S_t}{time+1}$;\\
time=time+1;\\
Broadcast $S$ and time;\\

}
\If{Receive a new $S$}
{Average the Received $S$ to $\overline{S}=\frac{\sum time\times S}{\sum time}$;\\
\If{$\|S-\overline{S}\|_p\leq D$}{Create a new block $Blk$ after longest chain;\\
	Broadcast $Blk$;\\
	 \Return $Blk$;\\}
}
\If{Receive a new $Blk$}
{
	\If{VerifyBlock($Blk$)==ture}
	{ Update $Blk$ to its chain;\\
		\Return $Blk$;
	}
}
\end{algorithm}

In Algorithm \ref{AlgShapley}, a miner first initializes the SV vector as an all-zero vector, and sets the calculation iteration number as $0$ (Line 1-2). The SV computation continues as long as one of the following two conditions are satisfied: 1) there is no new block received; or 2) the received block fails to pass the verification which is specified in Algorithm \ref{AlgVerification} (Line 3). The SV calculation process is described in Line 4 to Line 13. The miner initializes a temporary SV vector $S_t$ to record the calculated value in this iteration (Line 4). Then, the miner randomly generates a rank of the $K$ FL clients (Line 5). According to the rank, an SV of the first entity is calculated as in Line 6, which is the contribution of the entity to the loss function (Line 6)). For the next entity $i$, the Shapley value is calculated as its marginal contribution (Line 7-10). $S$ is updated by averaging all the previous iterations and the current $S_t$ (Line 11). The iteration time is then incremented by 1 (Line 12). Then, the entity broadcast $S$ and time (Line 13). 

\begin{figure}[ht]
	\centering
	\includegraphics[trim = 38mm 25mm 45mm 38mm,  clip, width=1.0\linewidth]{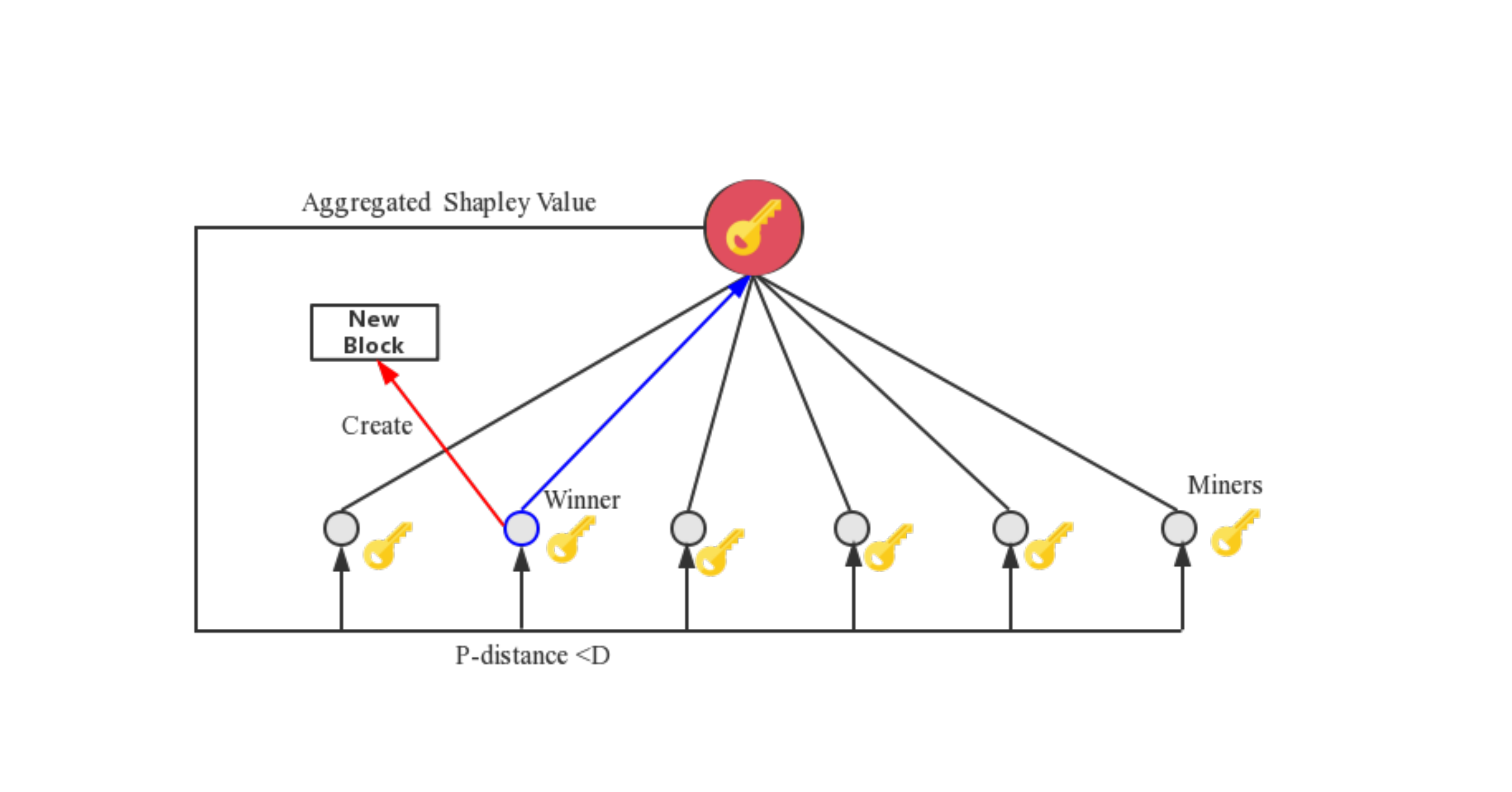}
	\caption{Shapley Valued based Consensus Protocol}
	\label{FigShapleyProof}
\end{figure}

Whenever a miner receives $S$ and $time$, the miner calculates the average results $\overline{S}$ of all the received $S$ (Line 16). Then, the miner calculates the $P$-distance between its own $S$ and $\overline{S}$. When the distance is no greater than the mining difficulty $D$, the miner becomes the winner and generates a new block $Blk$ (Line 18). The difficulty $D$ is adapted dynamically as explained in Section \ref{SectionDifficulty}. The illustration of the Shapley based verification is shown in Figure \ref{FigShapleyProof}. The new block is then appended to the current longest chain. 

Whenever a miner receives a new block $Blk$, the miner verifies this block according to Algorithm \ref{AlgVerification}. Once the verification passes, the block is appended to the miner's chain, and the mining process terminates (Line 23-28). 

\begin{figure}[!ht]
	\centering
	\includegraphics[trim = 44mm 25mm 18mm 30mm,  clip, width=1.0\linewidth]{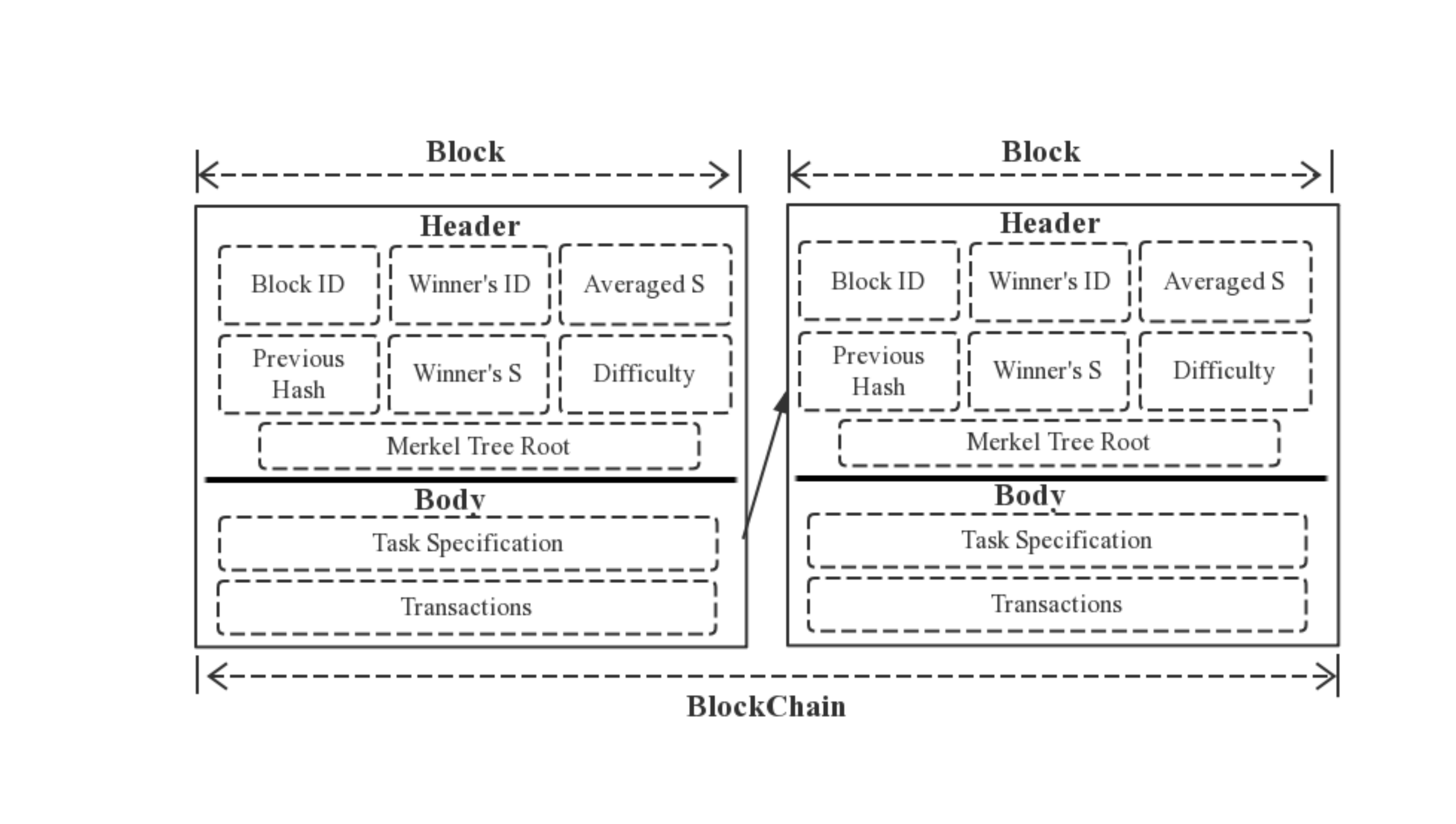}
	\caption{Block Structure in FedCoin}
	\label{FigBlock}
\end{figure}

The structure of a block is shown in Figure \ref{FigBlock}, including block header and block body. The block header includes seven pieces of information, Table \ref{TabBlock} presents the explanation about each header item. The block body records two types of data: (1) The task specification including all the inputs for Algorithm \ref{AlgShapley}; and (2) The transactions in the blockchain network. Here, a transaction is denoted as a certain amount of FedCoins transferred from a user account to another, which is similar to that in BitCoin \cite{nakamoto2008bitcoin}. The block winner has the privilege to create special transactions: transferring TrainPrice from its own account to to the FL clients according to $\overline{S}$. The detailed design is explained in Section \ref{SecIncentiveMechansim}.

\begin{table}[!htb]
	\caption{Explanation of Block Header}
	\label{TabBlock}
	\begin{tabular}{|l|p{5cm}|}
		\hline
		Type & Explanations\\
		\hline
		Block ID & The block height \\
		\hline
		Winner's ID & Identity of the block generator \\
		\hline
		Averaged $S$ & The calculated Shapley vector in Line 15 of Algorithm.\ref{AlgShapley}\\
		\hline
			Previous Hash & Hash value of previous block based on a hash function, e.g. SHA 256\\
		\hline
		Winner's $S$ & The Shapley vector calculated by the winner\\
		\hline
		Difficulty & The required difficulty $D$\\
		\hline
		Merkel Tree Root & Root of Merkel tree organized with transactions in block body\\
		\hline
	\end{tabular}
\end{table}

The verification procedure is described as in Algorithm \ref{AlgVerification}. Three conditions must be satisfied for a block to successfully pass the verification. The first condition is $\|S_t-\overline{S}_t\|_p\leq D$ which aims to verify whether the winner has generated the block with a valid SV calculation result. The second condition is $\|\overline{S}-\overline{S}_t\|_p\leq D$ which requires that the $\overline{S}$ value of the block should be close enough to the local aggregated $S$. $\overline{S}$ should be equal to $\overline{S}_t$ when the blockchain network is synchronized. In an asychronized network, this condition requires that the winner should aggregate a sufficient number of results from other entities. Thirdly, the current block ID should be the largest to ensure that only the longest chain is acceptable. This longest chain principle can effectively avoid forking, resulting in consistent chain status in a distributed network.  
\begin{algorithm}
	\caption{VerifyBlock (new Blk)}
	\label{AlgVerification}
	\KwIn{Blk: Received new block;\\
	\hspace{1.1cm}$\overline{S}$: Local average of received Shapley Value;\\
\hspace{1.1cm}$D$:  Difficulty in Mining;\\}
\KwOut{ValuationResult: True OR False}
	$S_t=Blk.S$; $\overline{S}_t=Blk.\overline{S}$;\\
	\If{$\|S_t-\overline{S}_t\|_p\leq D$}
	{
		\If{$\|\overline{S}-\overline{S}_t\|_p\leq D$}
		{
			\If{Blk.ID$\ge$ longest chain length}
			{\Return ValuationResult=ture;\\	
			}
		\Else{\Return ValuationResult=false;}
		}\Else{\Return ValuationResult=false;}
	}\Else{\Return ValuationResult=false;}
\end{algorithm}

\subsection{Dynamic Mining Difficulty}
\label{SectionDifficulty}
The difficulty level in mining new blocks can be adapted dynamically. There are two main factors influencing the difficulty updates: 1) the total mining power of the miners and 2) the speed of generating a block. Given the same mining power, the difficulty level should be decreased as the the block generation speed increases. Given the same block generation speed, the difficulty level should be increased as the mining power increases. Difficulty update can be achieved by deploying a smart contract. For example, in BitCoin, a block is generated in every ten minutes and the difficulty level is updated in every two-week duration. 

\subsection{The Payment Scheme}
\label{SecIncentiveMechansim}
With the FedCoin system in place, an FL model requester starts by depositing $V$ FedCoins in the FL server. The value of $V$ shall be no greater than the value of the FL model for the requester. To divide $V$ among FL clients, blockchain miners, and the FL server, all the entities should register a transaction account. The value of $V$ is then divided into three parts.
\begin{itemize}
\item TrainPrice: payments to the FL clients;
\item ComPrice: payment to the FL sever for processing the model aggregation;
\item SapPrice: payments to the blockchain network miners for calculating the Shapley value of each client. 
\end{itemize}

The division can be determined by a pre-agreed smart contract. For example, the division contract could specify that TrainPrice:ComPrice:SapPrice=7:1:2. Then, TrainPrice=0.7$V$, ComPrice=0.1$V$, and SapPrice=0.2$V$. The specific payment scheme is shown in Algorithm \ref{AlgIncentive}.

\begin{algorithm}[!htb]
	\caption{The Payment Scheme in FedCoin}
	\label{AlgIncentive}
	\KwIn{V: The value paid by a model requester;\\
		\hspace{1.1cm}$\overline{S}$: The final aggragated Shapley Value;\\}
	\KwOut{An allocation of $V$}
	\While{FL server receives $V$ from a model requester}
	{
		Calculate TrainPrice and SapPrice;\\
		Publish traing task with price TrainPrice;\\
		\If{The model is well trained}
		{Publish a Shapley task to blockchain network with pirce SapPrice;}
	}
	\While{a new block is mined}
	{
		FL server transfers TrainPrice+SapPrice to the block winner;
		\For{each FL client $i$}
		{
		\If{$S_i>0$}{$p_i=\frac{S_i}{\sum_{S_j>0}S_j}$TrainPrice;\\
			block winner transfers $p_i$ to $i$;}
	}
	}
\end{algorithm} 

In Algorithm \ref{AlgIncentive}, a model training task is successfully accepted by the FL server whenever the server receives payment $V$ from the FL model requester. The payment of $V$ is confirmed when the transfer transaction (requester$\xrightarrow{V}$server) is recorded in the blockchain. The server then calculates TrainPrice and SapPrice and leaving ComPrice=V-TrainPrice-SapPrice as its own payment for processing the task (Line 2). The training task is then published to FL clients with price TrainPrice (Line 3). When the training task is completed, the server then publishes a SV calculation task to the blockchain network with price SapPrice (Line 4-6). As the blockchain network completes the task by successfully mining a new block, the server creates a transaction to transfer TrainPrice+SapPrice to the block winner. The block winner creates the transactions in dividing TrainPrice to the FL clients with positive Shapley value. All the transactions as well as submitted unconfirmed transactions are stored in the new block. 

\section{Analysis}
Under FedCoin, the decentralized payment scheme is reliable based on the security of proposed PoSap consensus protocol. Each miner who successfully calculated the sufficiently converging SV is allowed to record a set of transactions and receive payment from the FL server. The more mining power (i.e. resources) a miner applies, the higher its chances to become a block winner. PoSap provides incentives for miners to contribute their resources to the system, and is essential to the decentralized nature of the proposed payment system. 

The security of PoSap is also similar to that of the BitCoin system. Empirical evidence shows that Bitcoin miners may form pools in order to decrease the variance among their incomes. Within such pools, all members contribute to the solution of each cryptopuzzle, and share the rewards proportionally to their contributions. Ideally, a feasible payment system should be designed to resist the formation of large mining pools. Bitcoin system has been shown to be vulnerable when a mining pool attracts more than 50\% of the miners. Similarly, the proposed system can also only resist upto 50\% of the miners colluding. 

Next we discuss how our system fares against the selfish mining strategy \cite{MajorityNotEnough2018}. 
\begin{observation}
When the FL server processes FL training model requests sequentially, it is not rational for colluders to follow the selfish mining strategy.
\end{observation}
According to Algorithm \ref{AlgIncentive}, each public block winner is paid by the FL server before creating a new block containing the block reward payment transactions. When the training task is processed one by one, if a selfish miner becomes the winner but does not publish this result immedietely, it cannot receive the block rewards. Meanwhile, the selfish miner cannot mine the next block without publishing its private block since the next SV task must wait for the completion of payment in the current block in the setting of sequentially training models.
\begin{observation}
When the FL server processes FL training model requests in parallel, and all the miners have the same block propogation delay to the FL server, the expected revenue for selfish miner is greater than that for honest miners when the selfish pool attracts more then 25\% of the total mining power in the blockchain network.
\end{observation}
If the tasks are published in parallel, a selfish miner can reserve a block and continue to mine the next block. The state transation and revene analysis is same as that in \cite{MajorityNotEnough2018}, resulting the condition of the threshold of selfish pool's mining power to be 1/4 under the condition of the same propogation delay to the FL server. Thus, processing FL training model requests in parallel under the proposed scheme is not recommended. 


\section{Experimental Evaluation}
To evaluate the proposed FedCoin, we set up a blockchain environment and design a federated learning task based on a real-world dataset. The objective of the experiments is to verify whether FedCoin can promote high quality data from distributed clients and whether the computational cost of PoSap is feasible.

\subsection{Experiment Settings}
We design our experiment based on a well-known digit classification dataset - MNIST with 70,000 images and a widely used software environment - TensorFlow - to perform federated digit classification tasks. 

We set up a FL server and 100 FL clients. The MNIST dataset is divided among the clients such that their data quality vary. We set there are 10 groups of 10 clients each. Clients from each group own datasets which belong to one of the 10 preset quality levels. We refer to this quality level as a client's \textit{type}. Each client of type $T_j$ is randomly assigned a training set following a uniform distribution over the $10-j$ class labels ($j=0,1,\cdots, 9$) as its local dataset. 
The FL training model in our experiments is classical neural network. We adopt the popular \texttt{FedAvg} as the FL aggragtion function, which averages the collected local model parameters to derive the global FL model parameters. Each local client trains the model for 20 iterations.  
\begin{figure*} [ht]
	\begin{minipage}[t]{0.31\linewidth} 
		\centering 
		\includegraphics[trim = 0mm 0mm 20mm 10mm,  clip, width=1\linewidth]{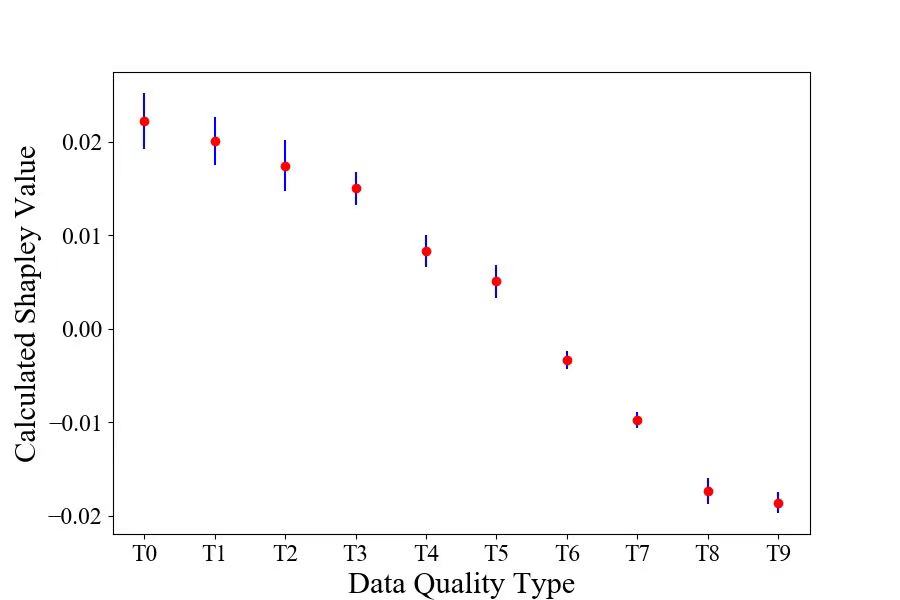}
		\caption{The Shapley Values Calculated in FedCoin for Clients in Different Types}
		\label{FigSV}
	\end{minipage}%
	\hspace{3mm}
	\begin{minipage}[t]{0.31\linewidth} 
		\centering 
		\includegraphics[trim = 0mm 0mm 20mm 10mm,  clip, width=1\linewidth]{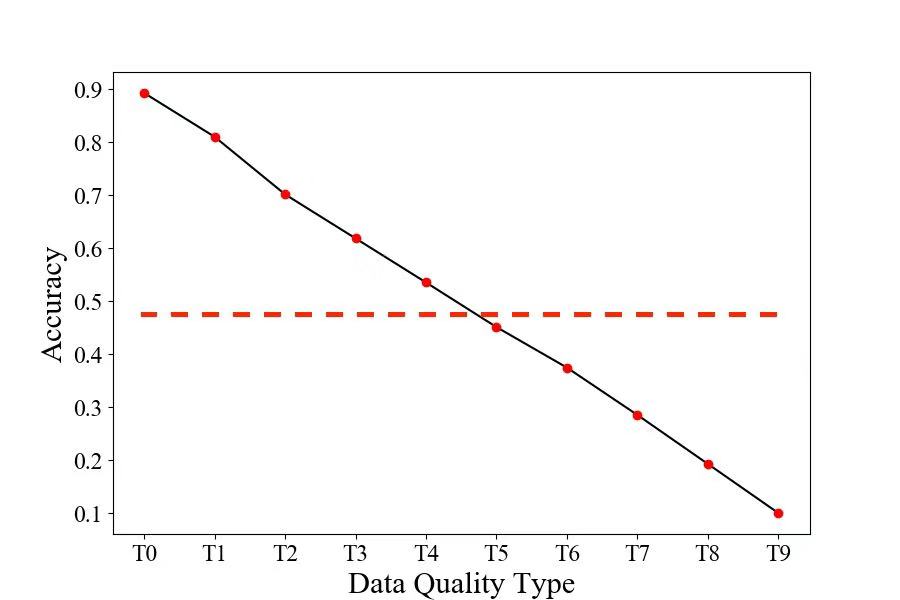}
		\caption{The Shapley Values Calculated in FedCoin for Clients in Different Types}
		\label{FigAccuracyEachType}
	\end{minipage} \hspace{3mm}
	\begin{minipage}[t]{0.35\linewidth} 
		\centering 
		\includegraphics[trim = 0mm 0mm 20mm 20mm,  clip, width=1\linewidth]{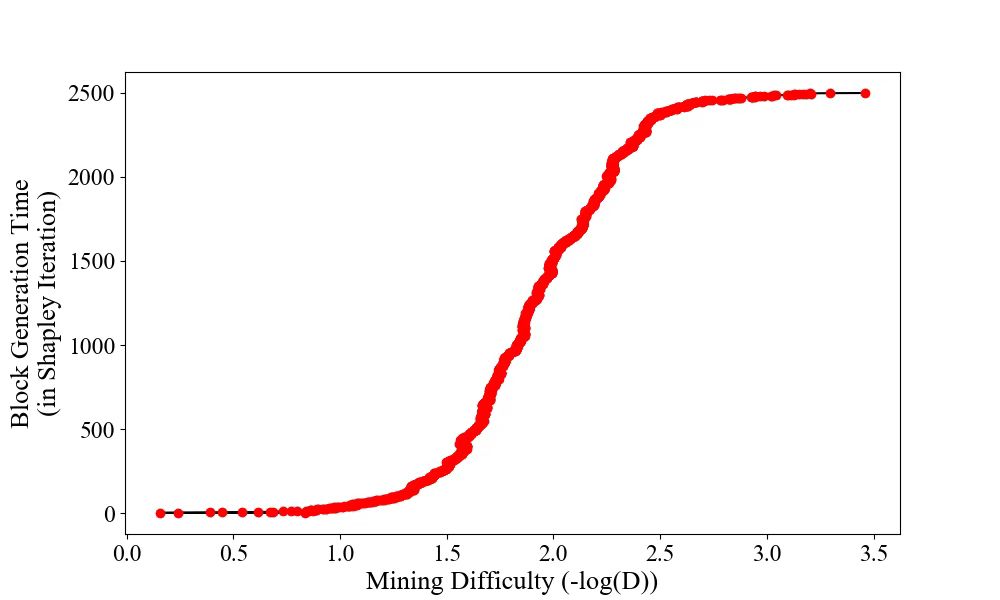}
		\caption{The Block Generation Time With Mining Varied Difficulty}
		\label{FigBlockGenDiff}
	\end{minipage}%
\end{figure*}

The consensus nodes are generated based on Docker. Each consensus node can independantly communicate with each other by sending messages, performing Shapley value calculation tasks following PoSap, and verifying blocks. The total computational power is equal to that of our simulation platform (CPU Interl i7-7700, GPU 2G, RAM 8g, ROM 1t, SSD 256M). We set $p=2$ (Euler distance) in PoSap for comparing mining difficulty.   


\subsection{Data Quality Evaluation}
We adopt Earth Mover's Distance (EMD) as a metric to measure the data quality from the perspective of data reliability in this experiment \cite{DBLP:journals/corr/abs-1806-00582}. EMD captures the distance for a client's training data distribution compared to a given distribution. Here, we use the distribution of the whole MINIST dataset as the comparison benchmark. A high EMD value for a given dataset indicates that the datast is of low quality. The data quality measured by EMD for each client type is shown in Table \ref{TabEMD}. We can observe that the data quality linearly decreases from 
$T_0$ to $T_9$
\begin{table}[!ht]
    \caption{Data quality of each client type.}
    \label{TabEMD}
    \centering
    \begin{tabular}{|c|c|c|c|c|c|}
    \hline
        Data Type &  $T_0$& $T_1$ &  $T_2$& $T_3$&  $T_4$\\
        \hline
        Quality (EMD) &  0& 0.02 &  0.04& 0.06&  0.08\\
        \hline
        Data Type & $T_5$ & $T_6$& $T_7$&  $T_8$& $T_9$\\
        \hline
        Quality (EMD) & 0.10&  0.12& 0.14& 0.16& 0.18 \\
        \hline
    \end{tabular}
\end{table}

\subsection{Results and Discussion} 
The average Shapley values calculated by FedCoin for the clients belonging to each type are shown in Figure \ref{FigSV}. The results in Figure \ref{FigSV} show that the proposed FedCoin approach can accurately compute clients' Shaply Values which are critical to incentive mechanisms designed for federated learning. We can also observe that the computed Shapley values of client type $T_0$ is higher than all the other clients with lower data quality. The Shapley values decrease with the quality level decays from $T_1$ to $T_9$. Moreover, only the values for type $T_0$ to $T_4$ are positive, indicating that only half of the clients can positively contribute model accuracy improvement. It also shows that our PoSap can effectively promote high quality data in collabratively FL application scenrio. The negative Shapley values for type $T_5$ to $T_9$ means the clients can mislead the model training.  

To empirically explain why there is negative Shapley values, we train the same model with clients from each type separately. The testing accuracy of the model trained by 10 clients in each type is presented in Figure \ref{FigAccuracyEachType}. It has been found that the model accuracy based data from clients with negative Shapley values is indeed lower than the accuracy achieved in the FL model based on the 100 clients. In our system, we will not reward the clients with negative Shapley values, discouraging these clients in misleading the model training.     


The block generation time with different mining difficulty values is shown in Figure \ref{FigBlockGenDiff}.
It is measured in the unit of Shapley calculation iterations, each of which executing Lines 3 to 14 in Algorithm \ref{AlgShapley} once. The x-axis contains the difficulty value in the form of $-log(D)$ (e.g., $x=2$ means $D=1e-2$). The y-axis contains the corresponding time required to generate a new block measured in Shapley calculation iterations executed by the block winner. It can be observed that the block generation time increases as the mining difficulty increases. As the mining difficulty increases beyond $1e-3$ ($D<1e-3$), the block generation time converges at about 2,500 Shapley calculation iterations. This shows that the computational cost for the consensus nodes is upper-bounded. In other words, the computational cost of PoSap is feasible.



\section{Conclusions}
In this paper, we propose FedCoin - a blockchain-based payment system to enable a federated learning system. It can mobilize free computational resources in the community to perform costly computing tasks required by FL incentive schemes. The Shapley value of each FL client, reflecting its contribution to the global FL model in a fair way, is calculated by the proof of Shapley (PoSap) consensus protocol. The proposed PoSap which replaces the traditional hash-based protocol in existing Bitcoin based blockchain systems. All the payments are recorded in the block in an immutable manner. Under FedCoin, paying out incentives to FL clients does not need to rely on a central FL server. 

Experimental results show that FedCoin is able to properly determine FL clients' Shapley Value-based contribution to the global FL model with an upper bound on the computational resource required for reaching consensus. To the best of our knowledge, FedCoin is the first attempt to leverage blockchain technology in federated learning incentive scheme research. Thereby, it opens up new opportunities for non-data owners to contribute to the development of the FL ecosystem \cite{FL:2019}.
\bibliographystyle{named}
\bibliography{FedChain}

\end{document}